\documentclass{aa}

\usepackage{graphics}
\usepackage{epsfig}
\usepackage{rotating}
\usepackage{graphicx}
\usepackage{color}
\usepackage{xspace}
\usepackage{txfonts}

\begin{document}
\color{black}
\title{An ISOCAM survey through gravitationally lensing galaxy clusters.
\thanks{Based on observations with ISO, an ESA project with instruments funded by ESA Member States (especially the PI
countries: France, Germany, the Netherlands and the United Kingdom) and with the participation of ISAS and
NASA}}

\subtitle{III. New results from mid-infrared observations of the cluster Abell 2219}
\authorrunning{Coia et~al.}
%\titlerunning{New results from mid-infrared observations of the cluster A2219}
\titlerunning{An ISOCAM survey through gravitationally lensing galaxy clusters. III.}
\author{
        D.~Coia \inst{1} \and
        L.~Metcalfe \inst{2, 3} \and
        B.~McBreen \inst{1}\and
        A.~Biviano \inst{4}\and
        I.~Smail \inst{5}\and
        B.~Altieri \inst{2, 3}\and
        J.-P.~Kneib \inst{6,7}\and
        S.~McBreen \inst{1,8}\and
        C.~Sanchez-Fernandez \inst{2} \and
        B.~O'Halloran \inst{9}
        }
\institute{     Department of Experimental Physics, University College, Belfield, Dublin 4, Ireland.
           \and XMM-Newton Science Operations Centre, European Space Agency, Villafranca del Castillo, P.O. Box 50727,
                28080 Madrid, Spain.
           \and ISO Data Centre, European Space Agency, Villafranca del Castillo, P.O. Box 50727, 28080 Madrid, Spain.
           \and INAF/Osservatorio Astronomico di Trieste, via G.B. Tiepolo 11, 34131, Trieste, Italy.
           \and Institute of Computational Cosmology, University of Durham, South Road, DH1 3LE, U.K.
           \and Observatoire Midi-Pyr\'{e}n\'{e}es, 14 avenue Edouard Belin, 31400
                Toulouse, France.
           \and California Institute of Technology, Pasadena, CA 91125, USA.
           \and Astrophysics Missions Division, Research and Scientific Support Department of ESA, ESTEC, Postbus 299, 2200 AG Noordwijk, The Netherlands.
           \and Dunsink Observatory, Castleknock, Dublin 15, Ireland.
} \offprints{D. Coia, \email{dcoia@bermuda.ucd.ie}}

\date{Received 7 July 2004 / Accepted 20 September 2004}

\abstract{ The massive cluster of galaxies Abell 2219 (z = 0.228) with two spectacular gravitational lensing
arcs was observed at 14.3\,$\mu$m (hereafter 15\,$\mu$m) with the Infrared Space Observatory and results were
published by Barvainis et al. (\cite{1999AJ....118..645B}). These observations have been reanalyzed using a
method specifically designed for the detection of faint sources that had been applied to other clusters. Five
new sources were detected and the resulting cumulative total of ten sources all have optical counterparts. The
mid-infrared sources are identified with three cluster members, three foreground galaxies, an Extremely Red
Object, a star and two galaxies of unknown redshift. The spectral energy distributions (SEDs) of the galaxies
are fit with models from a selection, using the program GRASIL. Best-fits are obtained, in general, with models
of galaxies with ongoing star formation. Infrared luminosities and star formation rates are obtained for six
sources: the cluster members and the foreground galaxies. For the three cluster members the infrared
luminosities derived from the model SEDs are between $\sim5.7\times10^{10}$\,$\mathrm{L}_{\odot}$ and $1.4\times
10^{11}$\,$\mathrm{L}_\odot$, corresponding to infrared star formation rates between 10 and 24
$\mathrm{M}_\odot~\mathrm{yr}^{-1}$. The two cluster galaxies that have optical classifications are in the
Butcher-Oemler region of the color-magnitude diagramme. The three foreground galaxies have infrared luminosities
between $1.5\times 10^{10}$\,$\mathrm{L}_\odot$ and $9.4\times 10^{10}$\,$\mathrm{L}_\odot$ yielding infrared
star formation rates between 3 and 16 $\mathrm{M}_\odot~\mathrm{yr}^{-1}$. Two of the foreground galaxies are
located in two foreground galaxy enhancements (Boschin et al. \cite{2004A&A...416..839B}). Including Abell 2219,
six distant clusters of galaxies have been mapped with ISOCAM and luminous infrared galaxies (LIRGs) have been
found in three of them. The presence of LIRGs in Abell 2219 strengthens the association between luminous
infrared galaxies in clusters and recent or ongoing cluster merger activity.  }

\maketitle

\section{Introduction}

The study of clusters of galaxies is of fundamental importance in understanding the mechanisms that drive galaxy
formation and evolution and in constraining cosmological theories for the formation of large-scale structures
(e.g. Bahcall \cite{1988ARA&A..26..631B}; Gladders et al. \cite{2002AJ....123....1G}; Schuecker et al.
\cite{2003A&A...402...53S}). Moreover clusters offer the opportunity of studying environmental effects on the
evolution of galaxies and, through the phenomenon of gravitational lensing, allow the observation of background
galaxies that would normally be below detection thresholds (e.g. Mellier \cite{1999ARA&A..37..127M}; Metcalfe et
al. \cite{2001IAUS..204..217M,2003A&A...407..791M}).

The comprehension of the properties of clusters of galaxies is growing rapidly thanks to a range of
exceptionally good observations taken using both ground-based (e.g. Yamada et al. \cite{2000PASJ...52...43Y}; 
B{\" o}hringer et al. \cite{2001A&A...369..826B}; Gladders et al. \cite{2002AJ....123....1G}; 
Andreon \& Cuillandre~\cite{2002ApJ...569..144A}) and space-based
telescopes, particularly in the infrared, optical and X-ray bands (e.g. Duc et al. \cite{2002A&A...382...60D}; 
Stanford et al. \cite{2002ApJS..142..153S}; Burke et al. \cite{2003MNRAS.341.1093B}).

Mid-infrared deep surveys with ESA's Infrared Space Observatory (ISO, Kessler et al. \cite{1996A&A...315L..27K})
have revealed a population of starburst field galaxies that evolve significantly with redshift and have a median
redshift of 0.8 (Aussel et al. \cite{1999A&A...342..313A}; Oliver et al. \cite{2000MNRAS.316..749O}; Serjeant et al.
\cite{2000MNRAS.316..768S}; Lari et al. \cite{2001MNRAS.325.1173L}; Elbaz et al. \cite{2002A&A...384..848E}; Gruppioni et
al. \cite{2002MNRAS.335..831G}; Metcalfe et al.
\cite{2003A&A...407..791M}). The mid-infrared emission from the ISOCAM sources accounts for most of the cosmic
infrared background (Altieri et al. \cite{1999A&A...343L..65A}; 
Franceschini et al. \cite{2001A&A...378....1F}; Elbaz \& Cesarsky \cite{2003Sci...300..270E}; 
Metcalfe et al \cite{2003A&A...407..791M}; Sato et al.
\cite{2003A&A...405..833S}). Mid-infrared data have been published for local clusters (e.g. Boselli et al.
\cite{1997A&A...327..522B,1998A&A...335...53B}; Contursi et al. \cite{2001A&A...365...11C}) and distant clusters
of galaxies (e.g. Pierre et al. \cite{1996A&A...315L.297P}; L{\' e}monon et al. \cite{1998A&A...334L..21L};
Soucail et al. \cite{1999A&A...343L..70S}; Fadda et al. \cite{2000A&A...361..827F}; Metcalfe et al.
\cite{2003A&A...407..791M}; Coia et al. \cite{coia}). The results of the observations revealed a wide dispersion
in the number of luminous infrared galaxies (LIRGs, Genzel \& Cesarsky \cite{2000ARA&A..38..761G}) in clusters.
New observations with the Spitzer Space Telescope (Rieke et al. \cite{2001AAS...199.4401R}; Dole et al.
\cite{2003ApJ...585..617D}; Werner et al. \cite{spitzer}) will further contribute to the understanding of the
infrared population of galaxies in clusters.

The cluster of galaxies Abell 2219 ($z = 0.228$) was observed using the 15\,$\mu$m (LW3 : 12 to 18 $\mu$m,
$\lambda_\mathrm{eff}= 14.3\,\mu m$) filter of the ISOCAM instrument on board ISO (Cesarsky et al.
\cite{1996A&A...315L..32C}). Results were published by Barvainis et al. (\cite{1999AJ....118..645B}, hereafter
BAH99) and consist of five sources with flux densities between 530 $\mu$Jy and 1100 $\mu$Jy.  One of the sources
was identified with the Extremely Red Object (ERO) \object{J164023+4644} at redshift $z\sim 1.048$ and the
remaining four sources were identified with galaxies of unknown redshift.

In this paper we present the results of our re-analysis of the ISOCAM observations of Abell 2219 using a method
specifically designed for the detection of faint sources (Metcalfe et al. \cite{2003A&A...407..791M}). Section
\ref{sec:cluster} contains a short description of the cluster. Data reduction, source extraction and photometric
calibration are described in section \ref{sec:obs}.  The results are given in section \ref{sec:results}.
Spectral energy distributions, infrared star formation rates and luminosities are in section \ref{sec:seds}. The
discussion is given in section \ref{sec:disc} and conclusions are summarized in section \ref{sec:concl}.

We adopt H$_0=70$\,km\,s$^{-1}$\,Mpc$^{-1}$, $\Omega_\lambda=0.7$ and $\Omega_\mathrm{m} =0.3$.  With this
cosmology, the luminosity distance to the cluster is $\mathrm{D} = 1138\,\mathrm{Mpc}$ and 1\arcsec\ on the sky
corresponds to 3.7 kpc.  The age of the Universe at the cluster redshift of 0.228 is 10.7 Gyr.

\section{The cluster\label{sec:cluster}}    %SECTION 2

\begin{figure*}  %figure 1 
\centering
%\resizebox{\textwidth}{!}{\includegraphics*{1648fig1.eps}}
\includegraphics[width=\textwidth,bb=-10 0 580 550]{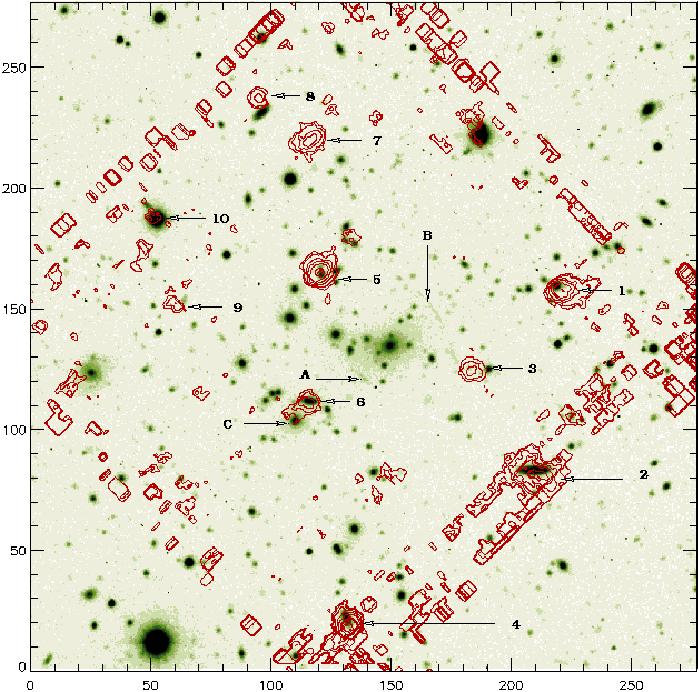}%}fig1.eps
\caption{Contours of the 15\,$\mu$m merged map overlaid on the B-band image of Abell 2219 (Smail et al.
\cite{1998MNRAS.293..124S}). The two gravitational lensing arcs are labelled with letters A and B and the second
brightest cluster member with the letter C (Smail et al. \cite{1995MNRAS.277....1S}). Sources published by BAH99
are labelled with numbers 1, 2, 4, 5, 7 (Table \ref{tab:res_tab}). The centre of the ISOCAM image is at
R.A.(J2000) 16$^\mathrm{h}$ 40$^\mathrm{m}$ 20.6$^\mathrm{s}$ DEC.(J2000) 46$^\mathrm{o}$ 42\arcmin\
39.6\arcsec. All axes are in arcseconds. North is up and East is to the left. \label{fig:over_iso}}
\end{figure*}

\begin{figure}%                                                    figure 2
  \resizebox{\hsize}{!}{\includegraphics{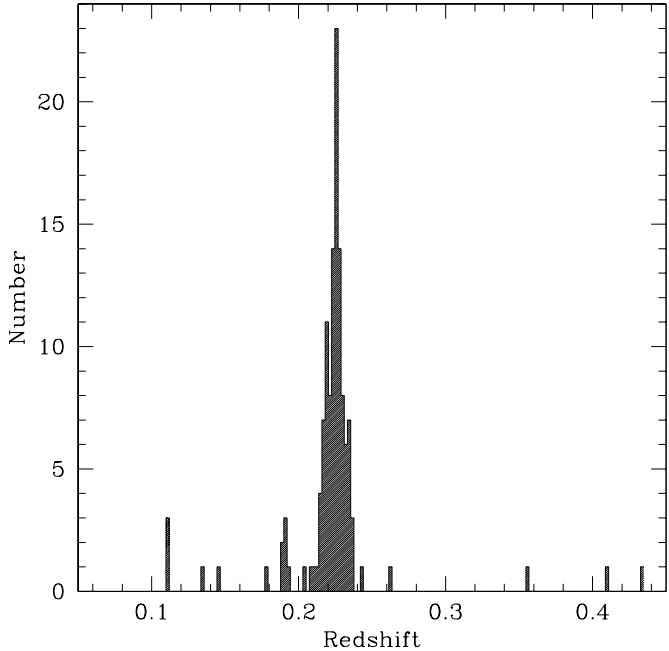}}
  \caption{The distribution of redshifts in the direction of the cluster Abell 2219 with two small
  foreground galaxy enhancements. Data are from Boschin et al. (\cite{2004A&A...416..839B}). }
\label{fig:boschin_fig2}
\end{figure}

\begin{table}[!t\textwidth]                      %    TABLE 1
\caption{Observational parameters: The observations were made with the ISOCAM LW3 filter (filter width: 12 -
18\,$\mu$m, $\lambda_\mathrm{eff}= 14.3\,\mu$m). On-chip integration time was always 10 seconds and the
6\arcsec~per-pixel-field-of-view was used. M and N are the number of steps along each dimension of the raster,
while dm and dn are the increments for a raster step. The table also includes the total area covered and the
number of readouts per raster step. Fifteen readouts were taken at the beginning of each of the observations to
allow for stabilization of the signal. The observations were repeated $k$ times, each time displacing the raster
centre by about 20 arcseconds.  \textsl{Tot. t} is the total observational time given by the sum of the
observation times of the three individual rasters. \label{tab:obs}}
    \begin{center}
     \begin{tabular}{ccccccccc}
               \hline
               \hline
\noalign{\smallskip}
$\lambda$&Reads   & \multicolumn{2}{c}{n Steps}&dm  &dn  & area        &Done k & Tot.t \\
($\mu$m)&per step& M          & N             &($\arcsec$)&($\arcsec$)&$(\arcmin^2)$ &times  & (sec)\\
\hline \noalign{\smallskip}
15.0&12        &3  &3                  &14  &14  &  13.4       &3      & 4074 \\
    \hline
      \end{tabular}
   \end{center}
\end{table}
Abell 2219 is a massive cluster with an estimated value for the mass of $2.8\times
10^{15}$\,$\mathrm{M}_\odot~\mathrm{yr}^{-1}$ and a virial radius of $\sim3.1$\,Mpc (Carlberg et al.
\cite{1997ApJ...476L...7C}; Boschin et al. \cite{2004A&A...416..839B}). It is an optically rich cluster with an
Abell richness class of 3 (Abell et al. \cite{1989ApJS...70....1A}). The core is dominated by a cD galaxy with a
minor concentration of galaxies visible around the second brightest cluster member at a projected distance of
190 kpc from the core.  The cluster exhibits two spectacular gravitational lensing arcs (Smail et al.
\cite{1995MNRAS.277....1S}), labelled A and B in Fig. \ref{fig:over_iso}. The distribution of redshifts of
cluster galaxies is shown in Fig. \ref{fig:boschin_fig2}.
Abell 2219 is one of the brightest X-ray clusters detected in the ROSAT All Sky Survey, with a luminosity in the
0.1-2.4 keV band of $1.8\times 10^{45}$\,$\mathrm{erg}$\,$\mathrm{s}^{-1}$ and a temperature, $\mathrm{T_X}$, of
$\sim 10$ keV (Allen et al. \cite{1992MNRAS.259...67A}; Rizza et al. \cite{1998MNRAS.301..328R}). The gas
distribution traced by the X-ray observations is elliptical in shape and centered on the cD galaxy. The major
axis of the ellipse is misaligned with respect to the major axis of the optical mass distribution, but lies
close to the axis defined by the cD galaxy and the second brightest cluster member, labelled C in Fig.
\ref{fig:over_iso}. This misalignment could indicate a past or ongoing merger of the cluster with the secondary
galaxy concentration detected in the optical (Smail et al. \cite{1995MNRAS.277....1S}; Boschin et al.
\cite{2004A&A...416..839B}).

Abell 2219 is characterized by a diffuse radio halo that extends for more than 2\,Mpc and is similar in shape to
the radio halo in Coma but 10 times more powerful (Bacchi et al. \cite{2003A&A...400..465B}). There may be a
connection between cluster mergers and the presence of extended and diffuse radio emission (Feretti
\cite{2003tsra.symp..209F}).

\section{Observations, data reduction, source detection and photometric calibration\label{sec:obs}}   % SECTION 3

Abell 2219 was observed with ISO in raster mode on April 20, 1997.  Three independent observations were made
using the LW3 ISOCAM filter. The parameters of the observations are given in Table~\ref{tab:obs}. The area
covered was 13.4 square arcminutes centered on the cluster core, corresponding to a physical area of
$0.8\times0.8\,\mathrm{Mpc}^2$. The diameter of the central maximum of the point spread function (PSF), at the
first Airy minimum, is $0.84\times\lambda(\mu m)$ arcseconds. The full-width at half-maximum (FWHM) is about
half that amount and Okumura (\cite{korio}) obtained a value of 4.9\arcsec\ at 15\,$\mu$m for the PSF FWHM in
the 6\arcsec~per pixel field-of-view.

The data were reduced using the software CIA (Delaney \& Ott \cite{ISOCAM}, Ott et~al.
\cite{1997adass...6...34O}) in conjunction with dedicated routines following the method described in Metcalfe
et~al. (\cite{2003A&A...407..791M}).  The three maps were merged into a single image thus increasing the
sensitivity to faint sources.  In the process the data were rebinned so that the final maps have a pixel size of
1\arcsec~ benefiting from the raster displacement and thus improving the resolution and allowing better
cross-identification with observations at other wavelengths. The 15\,$\mu$m merged map obtained is overlaid in
Fig.~\ref{fig:over_iso} on the Palomar B-band image of the cluster.

The extraction of sources was done using SExtractor (Bertin \& Arnouts \cite{1996A&AS..117..393B}).  The tool
was run separately on the three individual maps.  For the parameters chosen, the software could automatically
detect sources above $\sigma = 1.6$ over 9 consecutive pixels.  The program was then run on the merged image
with the same parameters used for the individual rasters.  A detection was considered real when present on the
merged map and on at least two of the three individual maps.  The source list obtained is given in
Table~\ref{tab:res_tab}.  The name of the ISOCAM source is derived from the satellite acronym (ISO), the partial
name of the cluster (A2219) and an index number of the 15\,$\mu$m source as listed in Table~\ref{tab:res_tab}
(e.g. ISO\_A2219\_01 is source 01 in Table~\ref{tab:res_tab}).  The index indicates increasing Right Ascension.

Montecarlo simulations involving the insertion of fake sources in the unprocessed data were performed on the
three individual rasters to analyse the effects of the data reduction process as described in Metcalfe et al.
(\cite{2003A&A...407..791M}).  The simulations were performed independently for all rasters and SExtractor was
applied with the same parameters adopted in the absence of the inserted fake sources.  The recovered signals for
the fake sources were typically 60-70\% of the inserted values.  The signals from the real sources were then
scaled according to the calibration factor determined from the simulations. The scaled signals were converted to
mJy using the filter specific ISOCAM calibration factor i.e. 1 ADU per gain per second $\mathrm{\equiv
0.51}$\,mJy for LW3 (Delaney \& Ott \cite{ISOCAM}).  A further scaling factor was applied to correct for
detector responsive transients effects. The steps for correcting for the detector responsive transients are
described in Metcalfe et al. (\cite{2003A&A...407..791M}), Section 4.3.  Taking the approach described there,
source signals recorded in the CAM LW3 filter in deep micro-scanned rasters were found to relate to stabilised
source signals (determined by applying the model of Coulais \& Abergel (\cite{2000A&AS..141..533C}) as
implemented in CIA) in the ratio: recorded\_signal/stabilised\_signal = $0.8(+0.05/-0.1)$.

The column named Precision in Table~\ref{tab:res_tab} is a measure of the accuracy of the photometric results.
The precision values are derived from fake source simulations. For each source brightness the precision is the
1-sigma scatter found in the recovered fluxes of fake sources of similar brightness inserted into the raw data.

\begin{table*}              % TABLE 2
\caption{List of 15\,$\mu$m sources in order of increasing R.A. From left to right: identification number;
source signal and precision in ADU; source flux-density and precision in mJy ; R.A. and DEC (J2000);
identification numbers and flux densities as given in BAH99; redshifts of the optical counterparts and source
identification numbers from the comprehensive optical catalogue of Boschin et al. (\cite{2004A&A...416..839B}).
The source ISO\_A2219\_05, marked with an asterisk, is assumed to be a member of the cluster.  The redshift of
source ISO\_A2219\_07 is from BAH99. The remaining redshifts are from Boschin et al.
(\cite{2004A&A...416..839B}). The flux densities of the five sources published by BAH99 are consistent with
those presented here. \label{tab:res_tab} } \centering
      \begin{tabular}{cccccccccccc}
               \hline
               \hline
\noalign{\smallskip}
ISO\_A2219&Flux  & Precision&Flux & Precision&\multicolumn{2}{c}{Coordinates}&\multicolumn{2}{c}{BAH99}&Redshift& Optical \\
  &(ADU) &(ADU)&(mJy)&  (mJy)   &R.A. (J2000)&  DEC. (J2000)& ID &Flux (mJy)  && ID\\
\hline \noalign{\smallskip}
01 &0.608 &0.029& 0.891&0.042   & 16 40 12.8 &  +46 43 04.0& 1&$0.890\pm0.110$ &0.230& 42\\
02 &0.673 &0.029& 0.985&0.042   & 16 40 14.0 &  +46 41 48.0& 2&$0.920\pm0.110$ &0.111& 45    \\
03 &0.289 &0.021& 0.433&0.030   & 16 40 16.4 &  +46 42 29.7&-&-&-&-\\
04 &1.072 &0.062& 1.558&0.090   & 16 40 21.4 &  +46 40 44.8&3&$1.420\pm0.110$&0.188&72  \\
05 &0.768 &0.055& 1.122&0.079   & 16 40 22.6 &  +46 43 12.0&4&$1.100\pm0.110$&$0.228^{\star}$&-\\
06 &0.313 &0.021& 0.468&0.030   & 16 40 23.1 &  +46 42 17.7&-&-&0.218&81  \\
07 &0.437 &0.029& 0.644&0.042   & 16 40 23.1 &  +46 44 05.1&5&$0.530\pm0.110$&1.048&-\\
08 &0.348 &0.021& 0.515&0.030   & 16 40 25.0 &  +46 44 23.2&-&-&0.135&90\\
09 &0.179 &0.021& 0.275&0.030   & 16 40 28.5 &  +46 42 57.8&-&-&-&-\\
10 &0.160 &0.021& 0.248&0.030   & 16 40 29.4 &  +46 43 32.5&-&-&-&Star\\
    \hline
      \end{tabular}
\end{table*}
\begin{figure*}[]      %figure 3
\begin{center}
\resizebox{0.75\textwidth}{!}{
\includegraphics{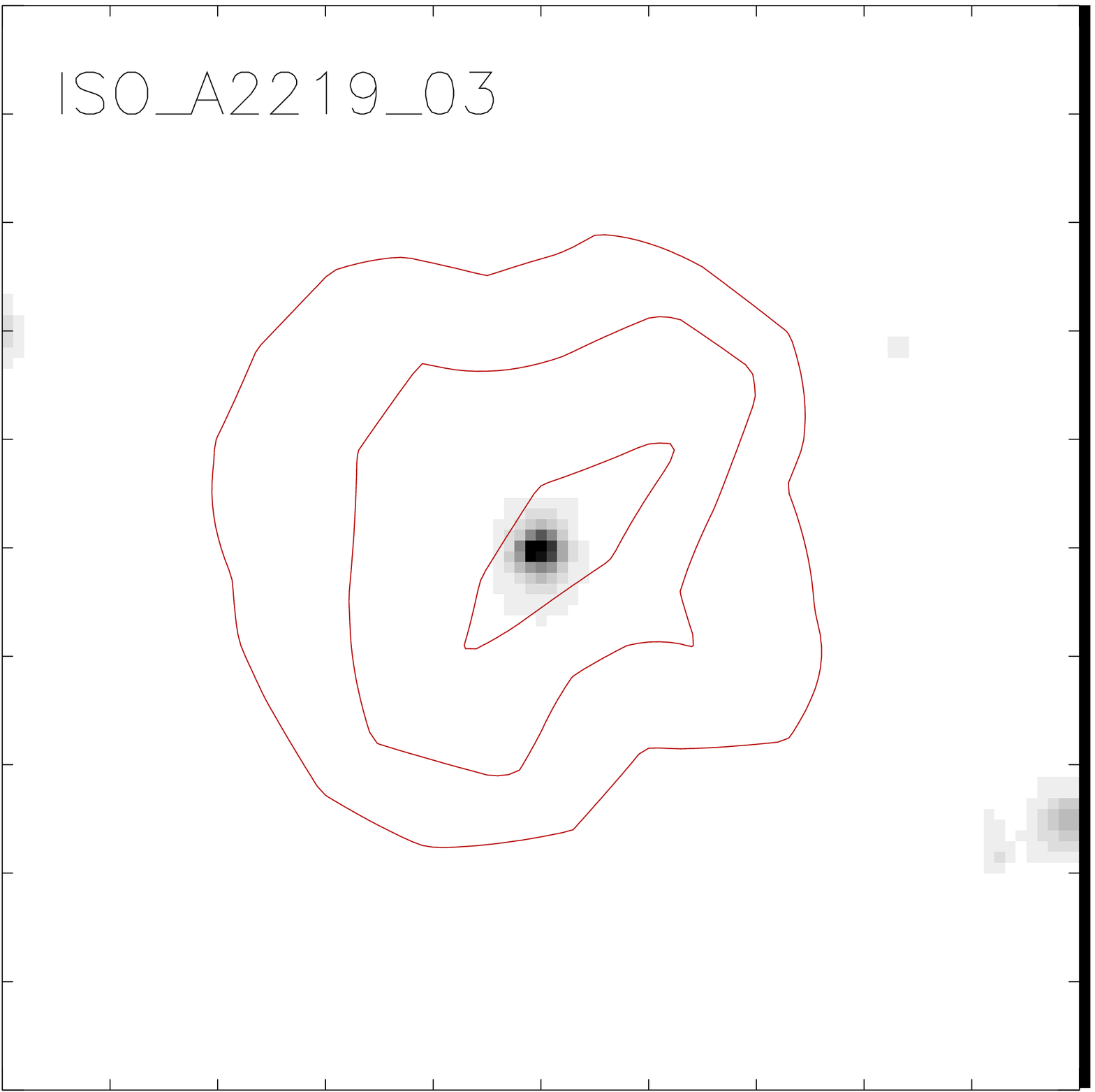}
\includegraphics{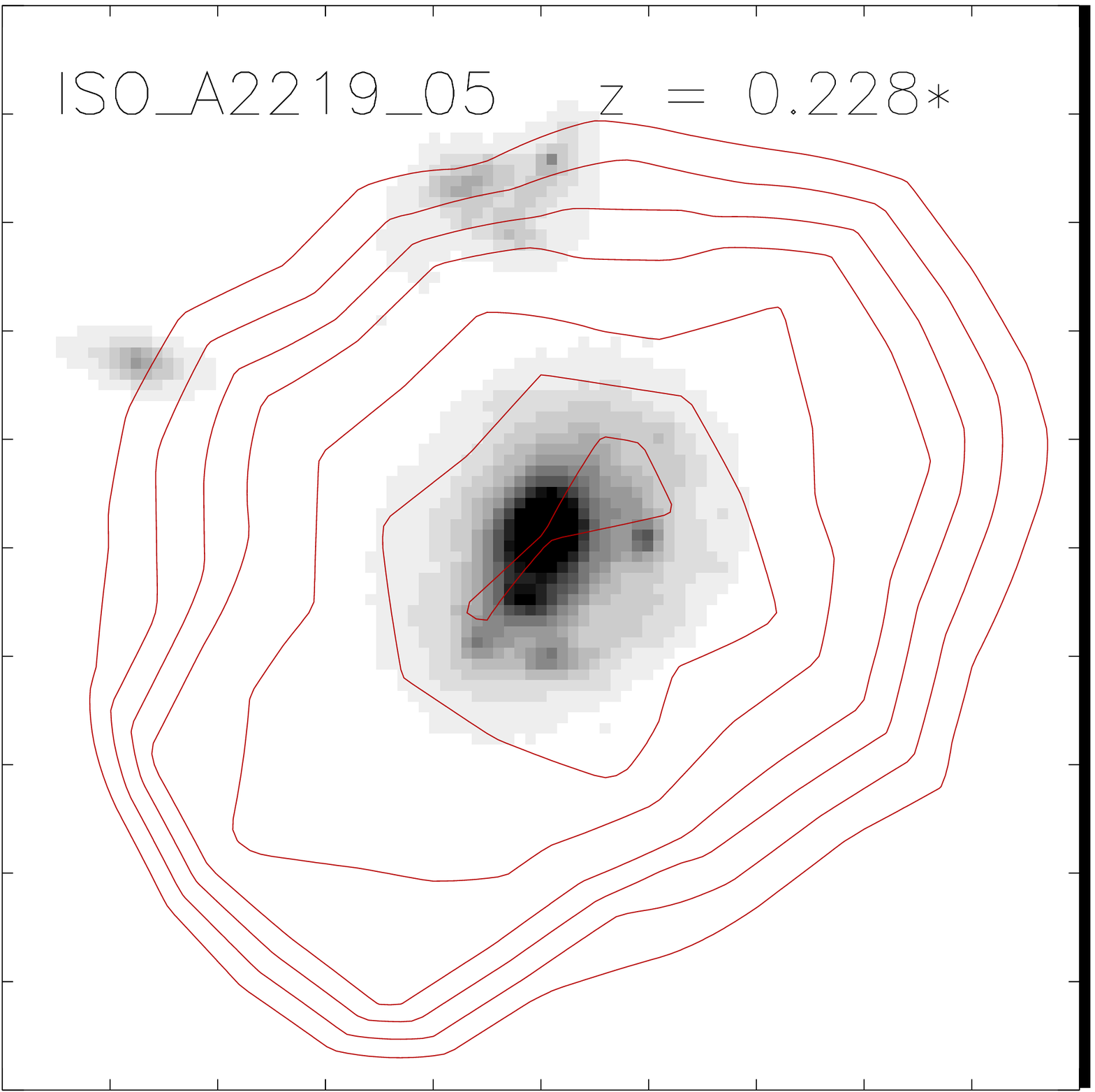}
\includegraphics{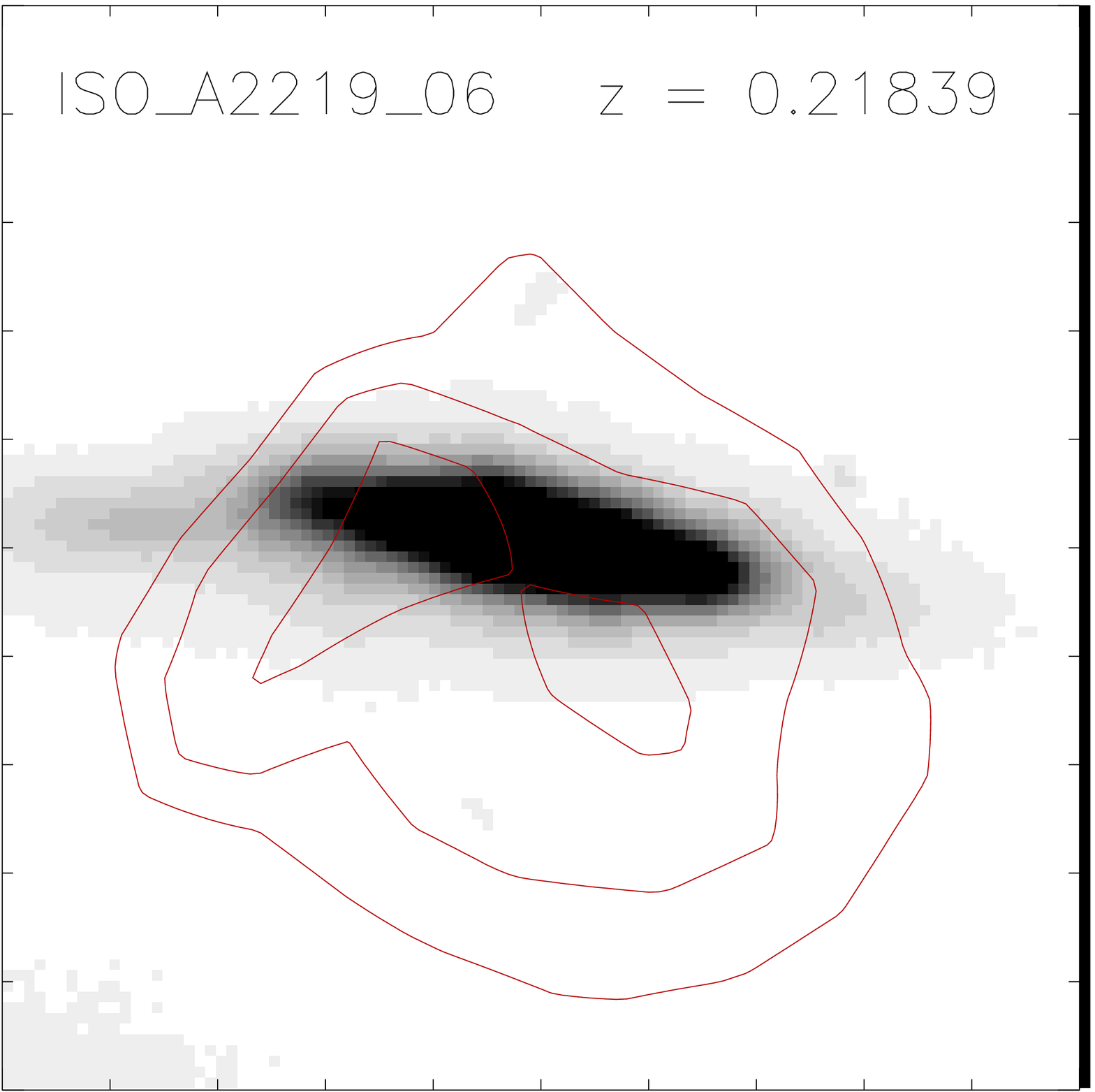}
\includegraphics{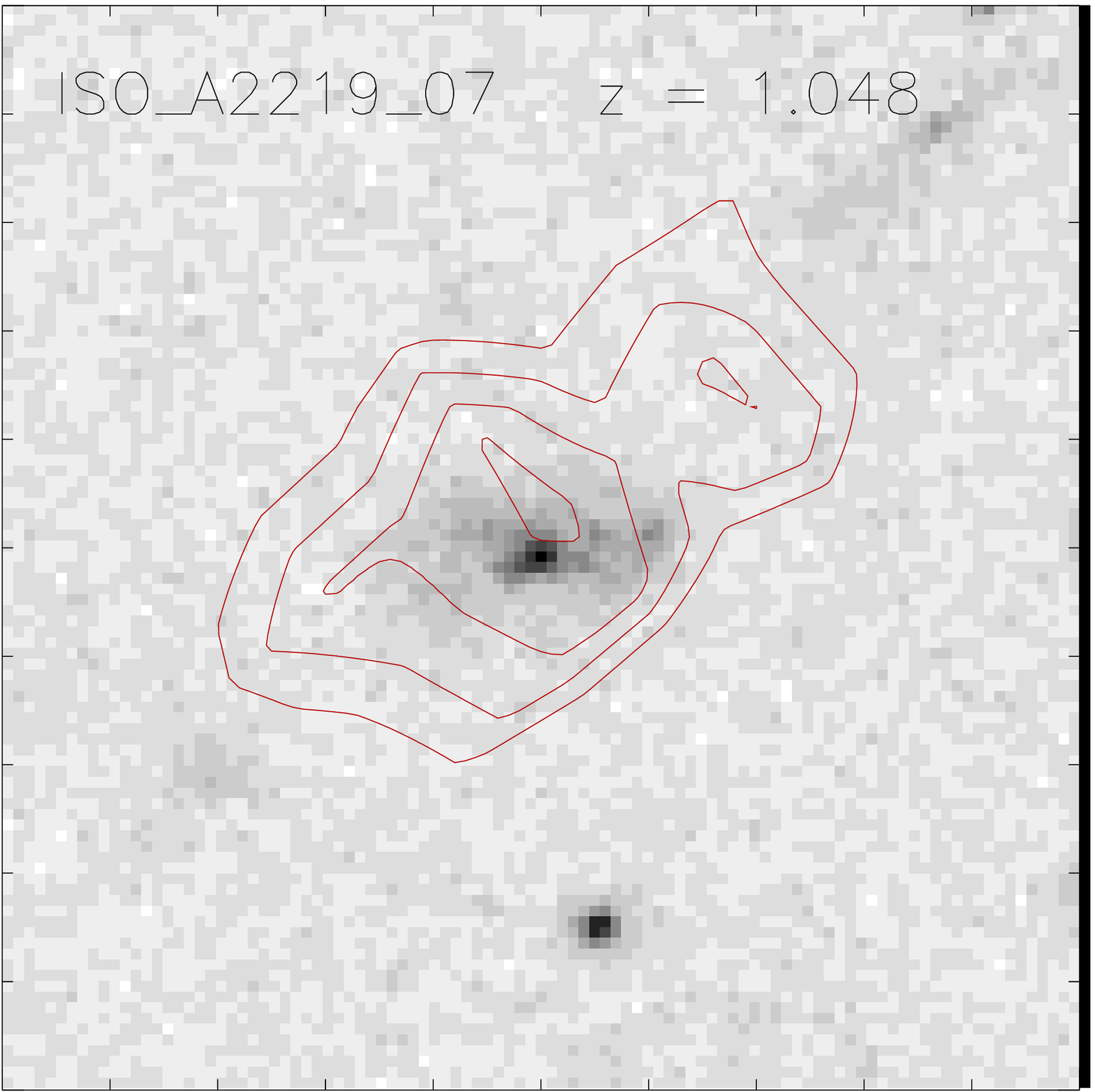}
} \caption{HST miniature maps of the four ISO 15\,$\mu$m sources within the HST field-of-view (Smith et al.
\cite{2002MNRAS.330....1S}). The name of the source is given in the top left hand corner of the figure and the
redshift in the top right hand corner. ISO\_A2219\_05 is assumed to be a member of the cluster, for reasons
mentioned in the text. All axes are in arcseconds. North is up and East to the left. \label{fig:wfpc}}
\end{center}
\end{figure*}
\begin{table*}
\centering \caption{Infrared luminosities and SFRs. The columns list: LW3 identification number as given in
Table \ref{tab:res_tab}; luminosity at 15\,$\mu$m; k-corrected total infrared luminosity; the k-corrections
derived as described in the text; infrared based SFRs; the model SEDs that give the best fit to the data;
alternative SED fits at the 95\% confidence; membership to the cluster (F: foreground galaxy; C: cluster
member). }\label{tab:sed_res}
 \leavevmode \begin{tabular}{cccccccccc} \hline \hline \noalign{\smallskip}
ISO\_A2219&L$_{15\mu \mathrm{m}}$&L$_\mathrm{IR}$& k-corr.&SFR[IR]                &Best-fit SED&other SED (95\%)& Membership \\
          & (L$_\odot$)          & (L$_\odot$)   &        &(M$_\odot$\,yr$^{-1}$) &  Model     &                &            \\
\hline \noalign{\smallskip}
01        & 2.20e+09             & 1.04e+11      & 1.2    & 18                    & S    (Y)   & -                      & C\\
02        & 3.92e+08             & 1.85e+10      & 1.0    & 3                     & SB+1 (O)   & S (O), SB+1 (Y)        & F \\
04        & 1.99e+09             & 9.36e+10      & 1.0    & 16                    & S    (Y)   & S (O), S (Y)          & F \\
05        & 3.00e+09             & 1.41e+11      & 1.4    & 24                    & SB+1 (O)   & S (O), S (O), SB+1 (Y)& C$^\star$\\
06        & 1.21e+09             & 5.73e+10      & 1.5    & 10                    & SB+1 (O)   & SB+1 (Y)               & C\\
08        & 3.15e+08             & 1.49e+10      & 1.0    & 3                     & SB+1 (O)   & S (O)                  &F\\
\noalign{\smallskip} \hline
\end{tabular}
\end{table*}

\section{Results\label{sec:results}}     % SECTION 4

The five sources published by BAH99 are confirmed.  In addition five new sources were found.  The sources are
listed in Table~\ref{tab:res_tab} and include the flux densities from BAH99.  The small differences in flux and
position between the present analysis and the results of BAH99 fall within the combined errors.

All of the 15\,$\mu$m sources have an optical counterpart (Fig. \ref{fig:over_iso}).  With the exception of
source ISO\_A2219\_10, which is associated with a star, all sources are identified with galaxies.  The sources
are compared with the spectroscopic catalog of Boschin et al. (\cite{2004A&A...416..839B}). ISO\_A2219\_01 and
ISO\_A2219\_06 are associated with the cluster galaxies 42 and 81 that are described as Butcher-Oemler galaxies
(Butcher \& Oemler \cite{1984ApJ...285..426B}), i.e. a description that refers to the excess of blue galaxies in
distant clusters relative to nearby clusters.  Galaxy 81 has weak x-ray emission indicating the presence of an
active nucleus. ISO\_A2219\_02 is identified with a galaxy in the foreground enhancement of galaxies at z
$\approx$ 0.11 and ISO\_A2219\_04 with another density enhancement at z $\approx$ 0.19 (Fig.
\ref{fig:boschin_fig2}).  They are listed in the catalog of Boschin et al. (\cite{2004A&A...416..839B}) as star
forming galaxies number 45 and 82. ISO\_A2219\_08 is listed in the same catalog as source 90. ISO\_A2219\_07 is
associated with ERO J164023+4644 at a redshift of $z = 1.048$. Sources ISO\_A2219\_03 and ISO\_A2219\_09 are
identified with galaxies of undetermined redshifts; their flux densities, as listed in Table \ref{tab:res_tab},
need to be corrected for gravitational lensing if they are confirmed as background sources, as suggested by the
faintness of the optical counterparts. ISO\_A2210\_05 is assumed to be in the cluster based on the dimensions of
its optical counterpart and location in a group of cluster galaxies. Moreover, it lies quite close to a mass
concentration in the lens model described in Smith et al. (\cite{astro-ph/0403588}), streghtening the conclusion 
that ISO\_A2210\_05 is indeed a cluster galaxy.

\subsection{Hubble Space Telescope images}      %SECTION 4.1

An area of 2.5\arcmin$\times$2.5\arcmin, centered on the cluster core, was observed by the Hubble Space
Telescope (HST) using the F702W filter on WFPC-2 (Smith et al. \cite{2002MNRAS.330....1S}).  The areas observed
with HST and ISOCAM partially overlap and the overlap region includes sources ISO\_A2219\_03, ISO\_A2219\_05,
ISO\_A2219\_06 and ISO\_A2219\_07.  For these sources, miniature maps centered on the ISOCAM coordinates were
extracted from the HST image and the 15\,$\mu$m contours are overlaid on the maps in Fig.~\ref{fig:wfpc}.

The HST image of ISO\_A2219\_03 shows a small face-on galaxy, probably an E/S0 galaxy since in this image there
is no evidence of a spiral structure.  However, the spatial resolution of the image is too poor to determine the
morphological type.  Given its angular dimensions ($\sim$4\arcsec$\cong$15\,kpc at the cluster redshift),
ISO\_A2219\_05 might be a cluster galaxy.  It shows disturbances at at least three sites in the outer regions
and an elongated bulge which could be the result of past or on-going mergers.  ISO\_A2219\_06 is associated with
an inclined spiral cluster galaxy.  The morphology of the ERO associated with ISO\_A2219\_07 appears irregular.
Its color and spectroscopic properties were thoroughly studied by Smith et al.
(\cite{2001ApJ...562..635S,2002MNRAS.330....1S}).

Overall, these four HST overlays indicate that the mid-infrared detections are not strongly associated with
galaxies of a specific morphological type, even though this sample is too small to draw definite conclusions.

\section{Spectral energy distributions, infrared star formation rates (SFRs) and luminosities\label{sec:seds}} %SECTION 5

Spectral energy distributions (SEDs) were computed for sources ISO\_A2219\_01, ISO\_A2219\_02, ISO\_A2219\_04,
ISO\_A2219\_05, ISO\_A2219\_06 and ISO\_A2219\_08 {using the program GRASIL (Silva et al.
\cite{1998ApJ...509..103S}) by including archival measurements in the optical and the near-infrared. Additional
photometric data were obtained by running SExtractor on optical and near-infrared images. These include images
taken in the B (Palomar Hale), I (WHT), J (WHT), and K (WHT) photometric bands (Smail et al.
\cite{1998MNRAS.293..124S}; Smith et al. \cite{2002MNRAS.330....1S}).
\begin{figure*}[t]                  %   FIGURE 4
\begin{center}
\includegraphics[height=2cm]{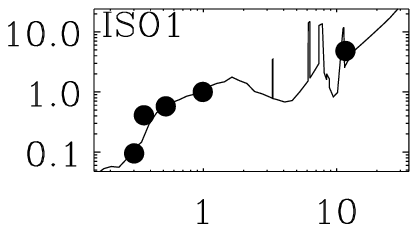}
\includegraphics[height=2cm]{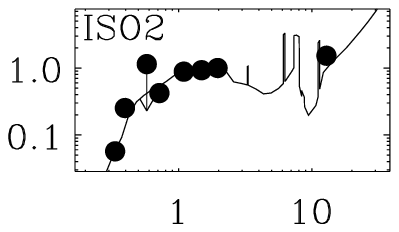}
\includegraphics[height=2cm]{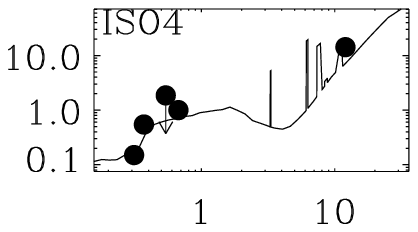}
\includegraphics[height=2cm]{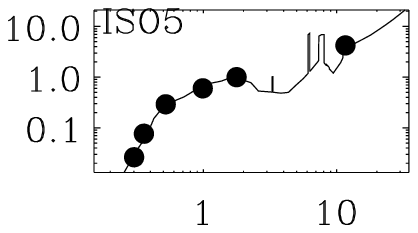}\\
\includegraphics[height=2cm]{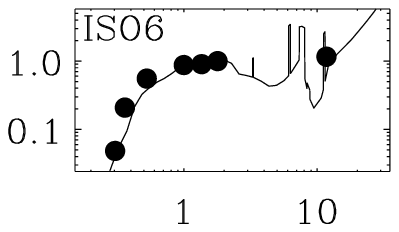}
\includegraphics[height=2cm]{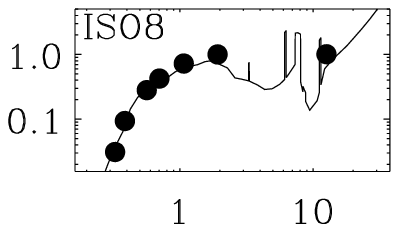}
\caption{The observations (dots) and model SEDs (continuous line) for sources ISO\_A2219\_01, ISO\_A2219\_02,
ISO\_A2219\_04, ISO\_A2219\_05, ISO\_A2219\_06 and ISO\_A2219\_08. The horizontal axis is the wavelength in the
cluster rest frame and the vertical axis is the flux density (in normalized units).} \label{fig:seds}
\end{center}
\end{figure*}

The observed data were compared with SED models that represent broad classes of spectral type, following a
procedure fully described in Coia et al. (\cite{coia}) and Biviano et al. (\cite{biviano}).

In this work, 20 models were considered. They were either taken from the public GRASIL library, or built by
running the publicly available GRASIL code, or kindly provided by L.~Silva (private comm.). The 20 models
reproduce the SEDs of several kinds of galaxies, from early-type, passively evolving ellipticals (labelled 'E',
in the following), to spiral galaxies (labelled 'S'), and starburst galaxies (similar to the local examples
Arp220 and M82), as seen either at the epoch of the starburst event, or 1 Gyr after (labelled 'SB' or 'SB+1',
respectively). Two SB models were considered, one involving as much as $\sim 10$\% of the total baryonic mass of
the galaxy, and another involving only $\sim 1$\% of the total baryonic mass of the galaxy. The strong burst
models generally provide a better fit to the SEDs of our galaxies.  At variance with Mann et al.
(\cite{2002MNRAS.332..549M}) we avoid considering models older than the age of the Universe at the cluster
redshift. Specifically, we consider models for galaxies with a formation redshift $z=1$, corresponding to an age
at the cluster redshift of $\sim 3$ Gyr, and models with a formation redshift $z=4$, corresponding to an age at
the cluster redshift of $\sim 8$ Gyr. We label the corresponding models with the suffixes 'Y' and 'O',
respectively.

The best-fitting SED model was determined by $\chi^2$ minimization, leaving the normalisation of the model as a
free parameter. The SEDs are plotted in Fig. \ref{fig:seds} and summarized in Table \ref{tab:sed_res}. S and
SB+1 models provide the SED best-fits and there seems to be no clear preference for Y or O models.

The k-corrections derived from the best-fit SEDs were used to obtain the rest-frame 15\,$\mu$m luminosities
using the relationship between the measured 15\,$\mu$m luminosities and the k-corrected total infrared
luminosities (L$_\mathrm{IR}$) from Elbaz et al. (\cite{2002A&A...384..848E}).

The source ISO\_A2219\_01 has a total infrared luminosity of $\sim 10^{11}$\,$\mathrm{L}_\odot$ (Table
\ref{tab:sed_res}) which classifies it as a LIRG. ISO\_A2219\_05 also has an infrared luminosity typical of a
LIRG, assuming it is a cluster member. ISO\_A2219\_06 has an infrared luminosity of $\sim6\times10^{10}$
$\mathrm{L}_\odot$, which is comparable to the luminosity of cluster sources found in \object{Abell 1689} by
Fadda et al. (\cite{2000A&A...361..827F}).

SFRs were computed from the mid-infrared fluxes and model SEDs following the conversion of Kennicutt
(\cite{1998ARA&A..36..189K}).
\begin{equation}\label{eq:kennicutt}
  \mathrm{SFR} \mathrm{[IR]} \simeq1.71\times 10^{-10}\ (\mathrm{L}_\mathrm{IR}/\mathrm{L}_\odot) \, \, \mathrm{M}_\odot\ \mathrm{yr}^{-1}
\end{equation}
The values obtained are listed in Table \ref{tab:sed_res}.  The infrared SFRs of the three cluster sources range
from 10 to 24 $\mathrm{M}_\odot~\mathrm{yr}^{-1}$, with a mean value of 17 $\mathrm{M}_\odot~\mathrm{yr}^{-1}$.
The infrared SFRs for the three foreground galaxies fall in the range between 3 and 16
$\mathrm{M}_\odot~\mathrm{yr}^{-1}$.  The high values of the SFRs in cluster and foreground galaxies are however
much less than the values $\approx 100 \,\, \mathrm{M}_\odot\ \mathrm{yr}^{-1}$ in field galaxies with redshifts
$\approx 0.8$ that were found in deep surveys with ISO (Oliver et al.
\cite{2000MNRAS.316..749O}; Mann et al. \cite{2002MNRAS.332..549M}; Elbaz \& Cesarsky \cite{2003Sci...300..270E}).

\section{Discussion} \label{sec:disc}

The dominant population of galaxies that fall in the Butcher-Oemler region of the color-magnitude diagramme of
Abell 2219 are a-type, i.e. galaxies with strong Balmer absorption ($\textrm{EW(H}\delta)>4\,\AA$, Boschin et
al. \cite{2004A&A...416..839B}).

ISO\_A2219\_01 and ISO\_A2219\_06 are identified with dusty star-forming galaxies 42 and 81 that are classified
as a-type with strong H$\delta$ absorption and weak [\ion{O}{ii}] emission with $\mathrm{EW}\sim 3\, \AA$
(Boschin et al. \cite{2004A&A...416..839B}).  The optical SFR for ISO\_A2219\_01 is estimated at $\sim 2$
$\mathrm{M}_\odot~\mathrm{yr}^{-1}$ which is much less than the value of 18 $\mathrm{M}_\odot\
\mathrm{yr}^{-1}$\, given in Table \ref{tab:sed_res}.  Most of the star formation is missed in the optical
because it is enshrouded by dust. In the clusters Abell 1689 (Fadda et al. \cite{2000A&A...361..827F}, Duc et
al. \cite{2002A&A...382...60D}) and \object{Cl 0024+1654} (Coia et al. \cite{coia}), that were observed with
ISO, the SFRs inferred using [\ion{O}{ii}] are about one-tenth of those calculated from the infrared
luminosities. The mid-infrared sources in Abell 1689 are detected in H$\alpha$, indicating that the star forming
processes are not completely hidden by dust (Balogh et al. \cite{2002MNRAS.335...10B}). The GRASIL models have
also been used to obtain the SEDs of the ISOCAM sources in Cl 0024+1654 (Coia et al. \cite{coia}) and
\object{Abell 2218} (Biviano et al. \cite{biviano}).  The galaxies in the Butcher-Oemler region of the
color-magnitude diagramme have SEDs that are best-fit by spiral models.  However the galaxies on the main
sequence in Cl 0024+1654 are usually best-fit with models of starburst galaxies as observed 1 Gyr after the
burst event.  A large number of cluster sources was detected by ISOCAM at 7~$\mu$m in Abell 2218. The SEDs of
most of these sources are best-fit by models of quiescent ellipticals with negligible SFRs.

A number of clusters with redshifts between $z = 0.18$ and $z = 0.39$ were mapped with ISOCAM and the numbers of
sources that have fluxes consistent with LIRGs, within the precision of the measurements, was determined for
each cluster.  A comparison was made between the number of LIRGs in the clusters taking into account virial
masses and radii, areas scanned and cluster distances (Coia et al. \cite{coia}). The cluster Cl 0024+1654 was
compared with the other clusters because it has a large number (10) of luminous sources that have measurements
consistent with LIRGs. This comparison revealed a large difference between Cl 0024+1654 ($z = 0.39$) and
\object{Abell 370} ($z = 0.37$) because in the latter only one source was detected whereas 8 were expected from
the comparison. The two clusters have similar mass, redshift and optical richness and the large difference in
the luminous infrared populations was attributed to a recent collision in Cl 0024+1654 (Czoske et al.
\cite{2002A&A...386...31C}) and the presence of gas rich progenitors to fuel the starbursts. The collision may
account for the large difference in the ratio of mass to infrared light for the two clusters.

The cluster Cl 0024+1654 was also compared with Abell 2218 ($z=0.18$), Abell 2390 ($z = 0.23$) and Abell 1689
($z = 0.18$) and a total of three sources was expected whereas none was detected. A similar comparison has now
been made between Cl 0024+1654 and Abell 2219.  In this case there is good agreement because only one source was
expected and two were detected implying that the clusters have similar values for the ratio of mass to infrared
light.

It is interesting that the redshift distributions along the lines-of-sight to the clusters Cl 0024+1654 and
Abell 2219 are quite similar. They have smaller foreground clusters that are displaced from the main cluster by
a small amount $\sim0.02$ in redshift (Fig. 2). In Cl 0024+1654 there is strong evidence that the two clusters
were recently involved in a collision (Czoske et al. \cite{2001A&A...372..391C,2002A&A...386...31C}). There is
also very strong evidence for merger activity in Abell 2219 as shown by the merging of small clumps of galaxies
with the cluster, the radio halo and powerful X-ray emission (Boschin et al. \cite{2004A&A...416..839B}). The
presence of LIRGs in Cl 0024+1654 and Abell 2219 strengthens the link between luminous infrared sources in
clusters with recent or ongoing merger activity.

The sample of clusters will be expanded by the ongoing observations with the Spitzer Space Telescope (Werner et
al. \cite{spitzer}).

\section{Conclusions\label{sec:concl}}      %SECTION 5

This paper presents new results from mid-infrared observations of the massive cluster Abell 2219.  The number of
mid-infrared sources is increased from 5 to 10 with respect to the earlier work of Barvainis et al
(\cite{1999AJ....118..645B}). All the sources have optical counterparts. Three sources are identified with
foreground galaxies, three with cluster galaxies, one with ERO J164023+4644 at $z \simeq 1.05$ and one with a
star.  The remaining two sources have undetermined redshifts and are likely to be background sources that are
gravitationally lensed by the cluster.

SEDs were obtained for the three cluster members and the three foreground galaxies. Model SEDs were used to
estimate parameters such as the infrared luminosity and SFR for the observed infrared sources. Models providing
the best-fits to the observed SEDs are those of actively star-forming galaxies, either massive spirals, or
starburst galaxies observed $\sim 1$ Gyr after the burst event. Two of the cluster galaxies are classified as
LIRGs and have a mean SFR of $\sim 17\,\mathrm{M}_\odot\,\mathrm{yr}^{-1}$, while the third has infrared
luminosity and SFR comparable to those star forming galaxies found in other clusters. The foreground galaxies
have infrared SFRs between 3 and 16 $\mathrm{M}_\odot~\mathrm{yr}^{-1}$ and two of them are located in
foreground galaxy enhancements.

The mid-infrared properties of the galaxies in Abell 2219 are compared with five distant clusters that were
mapped with ISOCAM.  The number of LIRGs detected in Abell 2219 agrees with the expected number when compared
with Cl 0024+1654 that is rich in luminous infrared galaxies.

\begin{acknowledgements}
JPK acknowledges support from CNRS  and Caltech.

The ISOCAM data presented in this paper were analyzed using CIA, a joint development by the ESA
Astrophysics Division and the ISOCAM Consortium.  The ISOCAM Consortium is led by the ISOCAM PI, C. Cesarsky. \\
This research has made use of data obtained from the High Energy Astrophysics Science Archive Research Center
(HEASARC), provided by NASA's Goddard Space Flight Center.
\end{acknowledgements}

%\listofobjects
\end{document}